\documentclass[conference]{IEEEtran}
\IEEEoverridecommandlockouts
\usepackage{multirow}
\usepackage{graphicx}
\usepackage{amssymb}
\usepackage{bmpsize}
\usepackage[cmex10]{amsmath}
\usepackage{algorithm}
\usepackage{algorithmic}
\usepackage{breqn}
\usepackage{amsmath}
\allowdisplaybreaks
\usepackage[utf8]{inputenc}
\usepackage{amssymb}
\usepackage{amsmath}
\usepackage{array}
\usepackage{eqparbox}
\usepackage[tight,footnotesize]{subfigure}
\usepackage{grffile}
\usepackage{multicol}
\usepackage{float}
\usepackage{url}
\usepackage{paralist}
\usepackage{multirow}
\usepackage{wrapfig}
\usepackage{color}
\usepackage{tikz}
\usetikzlibrary{shapes.geometric, arrows}
\usepackage{cite}
\usepackage{atbegshi}
\usepackage{setspace}
\usepackage{amsthm}

\newcommand{\RN}[1]{%
	\textup{\uppercase\expandafter{\romannumeral#1}}%
}
\def\BibTeX{{\rm B\kern-.05em{\sc i\kern-.025em b}\kern-.08em
    T\kern-.1667em\lower.7ex\hbox{E}\kern-.125emX}}
\begin{document}

\title{Sequential Processing of Observations in Human Decision-Making Systems}
\author{\IEEEauthorblockN{Nandan Sriranga{$^*$}, Baocheng Geng{$^{\dagger}$}, Pramod K. Varshney{$^*$}}
\IEEEauthorblockA{\kern-1em \kern-1em \kern-1em $^*$\text{Department of EECS}, \text{Syracuse University},  $^\dagger$\text{Department of Computer Science}, \text{University of Alabama at Birmingham}\\ 
\{nsrirang, varshney\}@syr.edu, \{bgeng\}@uab.edu}

}

\maketitle
\vspace{-0.3cm}
\begin{abstract}
In this work, we consider a binary hypothesis testing problem involving a group of human decision-makers. Due to the nature of human behavior, each human decision-maker observes the phenomenon of interest sequentially up to a random length of time. The humans use a belief model to accumulate the log-likelihood ratios until they cease observing the phenomenon. The belief model is used to characterize the perception of the human decision-maker towards observations at different instants of time, i.e., some decision-makers may assign greater importance to observations that were observed earlier, rather than later and vice-versa. The global decision-maker is a machine that fuses human decisions using the Chair-Varshney rule with different weights for the human decisions, where the weights are determined by the number of observations that were used by the humans to arrive at their respective decisions.               

\end{abstract}


\section{Introduction}
In various decision-making,  command and control systems operating in complex and uncertain environments,  incorporating human cognitive strengths and expertise is imperative to improve decision quality and enhance situational awareness. For example, in warning systems for natural disasters,  sensor measurements and human judgment are combined together for the early identification of hazards and risks. It is, therefore, imperative to study decision-making involving humans. 

The modeling and analysis of human decision-making in the
context of signal detection have become popular recently. There have been a few works that study human decision-making by incorporating factors related to human psychology in the statistical signal detection framework \cite{geng2019decision,9443353,Geng1911:Amelioration,8976222,9133140,sriranga2020human,gengheteroge,geng2021utility}. For example, the authors in \cite{geng2019decision} have investigated the impact of  random thresholds used by
human agents to make threshold-based decisions on the collaborative
detection performance. The Nobel prize-winning prospect theory \cite{tversky1992advances}, which provides a systematic formulation of judgment under biases,  has been employed to  model human decision-making behavior in the context of binary hypothesis testing \cite{Geng1911:Amelioration,9133140}. A semi-autonomous human-assisted decision-making framework where the human and the machine make correlated observations was proposed in \cite{sriranga2020human}. To help humans make better decisions, the optimal ordering of observations presented to humans under belief updating biases were studied in \cite{mourad2016real,mourad2018machine}.   Moreover, unlike a perfectly rational decision maker who always chooses the action that has the maximum utility,  it is possible for bounded rational humans to select all the actions in the action space but {better options are selected more often} \cite{anderson1992discrete}. The authors in \cite{9747866} employed
the bounded rationality model to quantify human uncertainty and evaluated individual decision-making performance when humans have different degrees of bounded rationality. 

In contrast \color{blue} with \color{black} the existing literature, our goal in this paper is to model and analyze how humans make decisions based on multiple observations \color{black} under belief updating biases and limited cognitive processing capabilities. As stated in \cite{anderson1981foundations}, information integration is a sequential process where information is received one piece at a time and integrated into a continuously evolving impression. Hence, we consider that local human decision-makers in a human team 
 observe a sequence of observations and they continuously update their beliefs until a decision is made. We investigate two types of cognitive biases and limitations that impact the human's decision quality: a) biases in belief updating, e.g., {\it conservatism} where the human is more adherent to prior knowledge and {\it over-responsiveness} where the human is more responsive to newly received information \cite{holt2009update} and, b) limited information processing capability in the sense that humans only use a limited number of observations in decision-making \cite{9413745}. The judgement biases and limitations may cause humans to behave quite differently from rational decision-makers, which may consequently impact the group decision-making performance.
\vspace{-0.1cm}

In this work, we first construct statistical models to characterize the above-mentioned cognitive biases and limitations in belief updating based on sequential observations. In particular, we use the weighted sum of log-likelihood ratios to model human belief updating biases and let a random variable represent the number of observations the human uses to make a decision. Next, we analyze the performance of individual human decision-makers in terms of the \textit{deflection coefficient}, using the expected values and variances of the individual humans' beliefs. Finally, we design the optimal decision fusion rule for collaborative human decision-making, given the individual behavioral properties of each human participant.

\section{Observation Model}
We consider a group of $N$ human decision-makers observing a sequence of observations (data streams) from the hypotheses 
\vspace{-0.07cm}
\begin{align}
\mathcal{H}_1 : X^{(t)}_{i}  \overset{\mathrm{iid}}{\sim}  f_1 \big(x^{(t )}_{i}\big), \quad 
\mathcal{H}_0 : X^{(t)}_{i}  \overset{\mathrm{iid}}{\sim}  f_0 \big(x^{(t)}_{i} \big). 
\label{eq:H0_obs}
\end{align}
\color{blue} The random variables $X^{(t)}_i$ represent the observations of the $i^{th}$ decision-maker at time-instant $t$. The observations $X_{i}(t)$ are independent and identically distributed (i.i.d) in time, conditioned on the hypothesis of origin and, the observations of any pair of decision-makers is also uncorrelated. \color{black}

\subsection{Human belief update}
The human decision-makers process the observations until a random time-instant $\tau^{*}_i$, where $\tau^{*}_i$ \color{blue} is a discrete random variable that may be generated according to a valid probability mass function (PMF). In this work, we consider the Geometric and Poisson RVs for the purpose of illustrating the results.\color{blue} The observations are processed for a duration of time $\tau_{i}^*$ and the belief update model employed by the human decision-makers is \color{black}
\vspace{-0.08cm}
\begin{align}
    \Lambda^{(t)}_i = \lambda^{(t)}_i + w_i \cdot \Lambda^{(t-1)}_{i} 
    \label{eq:belief_update}
\end{align}
\vspace{-0.08cm}
where $\lambda^{(t)}_i \equiv \lambda^{(t)}_i(x^{(t )}_{i}) = \text{ln} \Bigg ( \frac{f_1 \big(x^{(t )}_{i}\big)}{f_0 \big(x^{(t )}_{i}\big)} \Bigg )$ is the log-likelihood ratio of the most recent observation at time $t$, $\Lambda^{(t-1)}_{i} \equiv \Lambda^{(t-1)}_{i}(x^{(t-1)}_{i}, x^{(t-2)}_{i}, \dots, x^{(1)}_{i})$ is the cumulative belief until the previous time-instant $t-1$ and, $\Lambda^{(t)}_{i} \equiv \Lambda^{(t)}_{i}(x^{(t)}_{i}, x^{(t-1)}_{i}, \dots, x^{(1)}_{i})$ is the accumulated belief up to time instant $t$. 

The above formulation enables the mathematical modeling of the bias inherited due to belief updating. \color{blue} The parameter $w_i$ is a fixed (deterministic) parameter, which is an inherent property of the individual human decision-maker\color{black}. When $w_i$ is greater than 1, the decision maker attaches greater importance to observations that are more recent, whereas if $0< w_i < 1$, the observations that are observed earlier are assigned greater importance. With respect to the likelihood ratios, the effect can be seen such that at a time $t$, the likelihood ratio of the observation at a time $t-j$ is \textit{processed} as $\Bigg (\frac{f_{1}(x^{\tau^*_i}_{i})}{f_{0}(x^{\tau^*_i}_{i})} \Bigg)^{{w_i}^{t-j}}$, \color{blue} thereby biasing the likelihood-ratios of past observations\color{black}.

\section{Local Decision Rule}
{\color{blue}Upon} observing the data up to a time instant $\tau^{*}_{i}$, the human uses the accumulated belief until this time to make a decision 
\begin{equation}
    \Lambda^{(\tau^{*}_i)}_i \quad \underset{\mathcal{H}_1}{\overset{\mathcal{H}_0}{\lessgtr}} \quad \text{ln} \thinspace (T_i),
    \label{human_decision}
\end{equation}
which is in the form of a log-likelihood ratio test (LLRT), where $T_i$ is the decision threshold of the LLRT. This can be interpreted as a random sample-size LLRT, with the number of samples being a random function of the individual humans' decision-making behaviors. 

Unlike physical sensors and machines which are able to demonstrate consistent behavior as they are capable of being programmed into obeying explicit rules and commands, human decision-making is susceptible to inconsistencies due to the human's lack of motivation in fulfilling the task, fatigue, or impatience. All of these uncertain factors are jointly represented by the random stopping time $\tau^{*}_{i}$, which is a characteristic of the $i^{th}$ human decision-maker.      

The performance of the individual human decision-makers can be characterized by the quantities $\mathbb{E}[\Lambda_{i}^{(\tau^{*}_{i})}| \mathcal{H}_{1}]$, $\mathbb{E}[\Lambda^{(\tau^{*}_{i})}| \mathcal{H}_{0}]$, $\text{var}[\Lambda_{i}^{(\tau^{*}_{i})}| \mathcal{H}_{1}]$ and $\text{var}[\Lambda^{(\tau^{*}_{i})}| \mathcal{H}_{0}]$. 

The quantities $\mathbb{E}[\Lambda_i^{(\tau^{*}_i)}| \mathcal{H}_1]$ and $\mathbb{E}[\Lambda_i^{(\tau^{*}_i)}| \mathcal{H}_0]$ can be written as
\vspace{-0.1cm}
\begin{equation}
   \mathbb{E}[\Lambda_i^{(\tau^{*}_i)}| \mathcal{H}_k] = \mathbb{E} \Big[\lambda^{(j)}_i \Big \vert \mathcal{H}_k \Big] \cdot   \Bigg [ \frac{\mathbb{E}_{\tau^{*}_{i}}[w_{i}^{\tau_{i}^{*}}]-1}{w_i - 1}  \Bigg ].
   \label{eq:expected_LLR_belief_main}
\end{equation}
where $k \in \{0,1\}$ and \color{blue} $\lambda^{(j)}_i$ is the log-likelihood ratio of the $i^{th}$ human decision-maker at any arbitrary time-instant $j< \tau^{*}_{i}$, as the observations of any human decision-maker are identically distributed in time, according to (\ref{eq:H0_obs}) \color{black}. Further, the second term in the product in (\ref{eq:expected_LLR_belief_main}) can be written as
\vspace{-0.08cm}
\begin{align}
     \frac{\mathbb{E}_{\tau^{*}_{i}}[w_{i}^{\tau_{i}^{*}}]-1}{w_i - 1}  = 
     \begin{cases}
        \mathbb{E}_{\tau^{*}_{i}}[\tau_{i}^{*}]  & \text{if } w_i = 1\\
        \frac{\mathbb{E}_{\tau^{*}_{i}}\Big[e^{ \text{ln}(w_i) \thinspace \tau_{i}^{*}} \Big]-1}{w_i - 1} & \text{otherwise},  
    \end{cases}
    \label{eq:expected_time_RV_main}
\end{align}
which can be obtained by using the moment-generating function (MGF) of the RV $\tau^{*}_{i}$. Note that the quantity in (\ref{eq:expected_time_RV_main}) is always greater than or equal to 1 (obtained by using Jensen's inequality), which indicates that when the human decision-maker uses more than one observation, in the average sense, the expected value of the cumulative belief is larger than that with just one observation, $\mathbb{E} \Big[\lambda^{(j)}_i \Big \vert \mathcal{H}_k \Big]$. 

The quantities $\text{var}[\Lambda_i^{(\tau^{*}_i)}| \mathcal{H}_1]$ and $\text{var}[\Lambda_i^{(\tau^{*}_i)}| \mathcal{H}_0]$ can be written as
\vspace{-0.08cm}
\begin{align}
    \text{var}[\Lambda_i^{(\tau^{*}_i)}| \mathcal{H}_k] = &  \quad \text{var} \Big[\lambda^{(j)}_i \Big \vert \mathcal{H}_k \Big] \cdot   \Bigg [ \frac{\mathbb{E}_{\tau^{*}_{i}}[w_{i}^{2 \cdot \tau_{i}^{*}}]-1}{w^2_i - 1}  \Bigg ] \nonumber\\
    &+ \mathbb{E}\Big [\lambda^{(j)}_i \bigg \vert \mathcal{H}_k \Big] \cdot  \Bigg [ \frac{\text{var}_{\tau^{*}_{i}}[w_{i}^{\tau_{i}^{*}}]}{(w_i - 1)^2} \Bigg ] 
    \label{eq:var_LLR_belief_main}
\end{align}

The derivations for the expressions in (\ref{eq:expected_LLR_belief_main}) and (\ref{eq:var_LLR_belief_main}) are omitted due to space limitations but are available in \underline{https://arxiv.org/pdf/2301.07767.pdf}. The expressions are obtained by a straightforward application of the laws of \textit{total expectation} and \textit{total variance} respectively, on the individual humans' belief LLR.  

Further, the expressions for the overall probabilities of detection $P_{D,i}$ and false-alarm $P_{FA,i}$ for the human decision-maker are 
\vspace{-0.1cm}
\begin{align}
    P_{D,i} &= \sum_{j=1}^{\infty} \text{Pr}(\tau_{i}^{*} = j) \cdot \text{Pr}( \Lambda^{(\tau^{*}_i)}_i > \text{ln}(T_i)| \thinspace \thinspace \tau^{*}_{i} = j, \mathcal{H}_1) \nonumber\\
    &= \sum_{j=1}^{\infty} \text{Pr}(\tau_{i}^{*} = j) \cdot P_{D,i}(\tau^{*}_{i}) \quad \text{and},
    \label{eq:prob_det_human}
\end{align}
\vspace{-0.1cm}
\begin{align}
    \hspace{-2.9cm} P_{FA,i} = \sum_{j=1}^{\infty} \text{Pr}(\tau_{i}^{*} = j) \cdot P_{FA,i}(\tau^{*}_{i}).
    \label{eq:prob_fa_human}
\end{align}
The expressions for the quantities in (\ref{eq:prob_det_human}) and (\ref{eq:prob_fa_human}) are difficult to compute in closed-form due to which we use the deflection coefficient of the human belief LLR, \color{black} as a surrogate metric to characterize the detection performance of human decision-makers in this work. The metric can be computed using the expected value and the variance of the beliefs in equations (\ref{eq:expected_LLR_belief_main}) and (\ref{eq:var_LLR_belief_main}) and is defined as 

\begin{equation}
    \Delta_{i, k} = \frac{\Big(\mathbb{E}[\Lambda_{i}^{(\tau^{*}_{i})}| \mathcal{H}_{1}] - \mathbb{E}[\Lambda_{i}^{(\tau^{*}_{i} )}| \mathcal{H}_{0}] \Big)^2}{\text{var}[\Lambda_{i}^{(\tau^{*}_{i})}| \mathcal{H}_{k}]}, 
    \label{eq:deflection_coefficient}
\end{equation}
where $k=0,1$ and reflects the variance of the human belief. 

\begin{figure}[!h]
    \centering
    \includegraphics[width=60 mm]{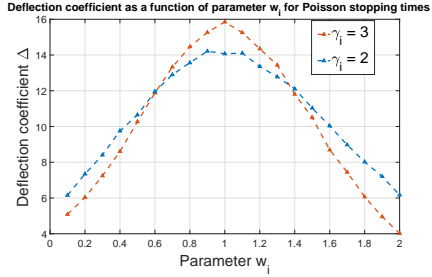}
    \caption{Deflection coefficient for different values of $w_i$, for different Poisson stopping times ($\tau^{*}_{i} \sim \text{Poisson}(\gamma_i)$). } 
    \label{fig:def_coeff_poisson}
\end{figure}

We illustrate the nature of the deflection coefficient as a function of the human decision-maker's parameter $w_i$ in Fig. \ref{fig:def_coeff_poisson}. When the human decision-maker observes data from $\mathcal{N}(s, \sigma^2)$ and $\mathcal{N}(0, \sigma^2)$ under the alternative ($\mathcal{H}_1$) and null ($\mathcal{H}_0$) hypotheses respectively. We consider $s = 5$ and $\sigma^2 = 1$ for this example. 

It can be seen that the human decision-maker performs well in terms of the deflection coefficient of their belief LLR when $w_i = 1$, i.e when the observations are processed rationally in comparison with values of $w_i \neq 1$. This is due to the fact that when $w_i<1$, $P_{D,i}$ decreases as $w_i \to 0$ since the LLRT neglects the information from earlier observations whereas when $w_i > 1$, the value of $P_{D,i}$ increases but at the cost of an increase in the value of $P_{FA,i}$. 

\section{Global Decision Rule}
In order to fuse individual human decisions, the Chair-Varshney rule \cite{ChairV1986} is employed at the FC. The human decisions which are denoted by $u_i \in \{-1, +1\}$ for $i=1, \dots, N$, where $u_i = 1$ when $\mathcal{H}_1$ is declared by the human decision-maker and $u_i = -1$ when $\mathcal{H}_0$ is declared, are combined at the FC as follows
\vspace{-0.1cm}
\begin{equation}
    \sum_{\{i: \thinspace u_i = 1\}}   \text{ln}\Bigg(\frac{P_{D,i}}{P_{FA,i}}\Bigg) + \sum_{\{i: \thinspace u_i = -1\}} \text{ln}\Bigg(\frac{1-P_{D,i}}{1-P_{FA,i}}\Bigg) \underset{\mathcal{H}_1}{\overset{\mathcal{H}_0}{\lessgtr}} \thinspace \thinspace T_g,
    \label{global_decision_rule}
\end{equation}
where $T_g$ is the threshold employed at the FC to yield the global decision.

When the FC has knowledge of the realizations of the number of observations $\tau_i^{*}$ used by each decision maker to arrive at their respective decisions $u_i$, the fusion rule can be simplified in the following way:
\vspace{-0.1cm}
\begin{align}
    &\prod_{i=1}^{N} \frac{p(u_i, \tau_i^{*}| \mathcal{H}_1)}{p(u_i, \tau_i^{*}| \mathcal{H}_0)} \quad \underset{\mathcal{H}_1}{\overset{\mathcal{H}_0}{\lessgtr}} \quad t_g \nonumber \\
   \implies &\prod_{i=1}^{N} \frac{p(u_i|\tau^{*}_{i}, \mathcal{H}_1) \cdot p(\tau^{*}_{i})}{p(u_i|\tau^{*}_{i}, \mathcal{H}_0)\cdot p(\tau^{*}_{i})} \quad \underset{\mathcal{H}_1}{\overset{\mathcal{H}_0}{\lessgtr}} \quad t_g  \nonumber \\
   \implies &\prod_{i=1}^{N} \frac{p(u_i|\tau^{*}_{i}, \mathcal{H}_1)}{p(u_i|\tau^{*}_{i}, \mathcal{H}_0)} \quad \underset{\mathcal{H}_1}{\overset{\mathcal{H}_0}{\lessgtr}} \quad t_g .
\end{align}
By taking the logarithm on both sides and separating the terms in the \color{blue} summation \color{black} based on whether $u_i = 1$ or $u_i = -1$, the global decision rule can be simplified to
\vspace{-0.1cm}
\begin{align}
    \sum_{\{i: \thinspace u_i = 1\}} &  \text{ln}\Bigg(\frac{P_{D,i}(\tau_i^{*})}{P_{FA,i}(\tau_i^{*})}\Bigg)  \nonumber\\
    &+ \sum_{\{i: \thinspace u_i = -1\}} \text{ln}\Bigg(\frac{1-P_{D,i}(\tau_i^{*})}{1-P_{FA,i}(\tau_i^{*})}\Bigg) \quad \underset{\mathcal{H}_1}{\overset{\mathcal{H}_0}{\lessgtr}} \thinspace \thinspace T_g,
    \label{global_decision_rule_simplified}
\end{align}
where $P_{D,i}(\tau_i^{*})$ and $P_{FA,i}(\tau_i^{*})$ are the probabilities of detection and false-alarm respectively, given that the number of observations used by the $i^{th}$ human decision-maker is $\tau_i^{*}$ and, $T_g = \text{ln}(t_g)$.

The fusion rule in (\ref{global_decision_rule_simplified}) is different from the rule in (\ref{global_decision_rule}), in the sense that the weights used for the different human decision-makers vary with each instance of the test at the FC due to the fact that each human decision-maker conducts an LLRT with a random number of samples. \color{blue} To perform \color{black} the decision rule in (\ref{global_decision_rule}), \color{blue} the FC \color{black} does not require the knowledge of the parameter $w_i$, the test threshold $T_i$ used by the humans or the number of observations used to arrive at a decision $\tau^{*}_{i}$, \color{blue} as long as $P_{D,i}$ and $P_{FA,i}$ in Equations (\ref{eq:prob_det_human}) and (\ref{eq:prob_fa_human}) are known \color{black}. The fusion rule in (\ref{global_decision_rule_simplified}) requires knowledge of these parameters but the weights used for the rule are analytically tractable. \color{black}

The weights, which are determined by $P_{D,i}(\tau^{*}_{i})$ and $P_{FA,i}(\tau^{*}_{i})$, are easier to compute in this case in comparison with the overall probabilities of detection and false-alarms $P_{D,i}$ and $P_{FA,i}$, as they involve the probabilities of the sum of a known number of random variables exceeding some value (one minus the CDF of a random variable or the Q-function for Gaussian RVs). Specifically, when the observations under the two hypotheses are distributed as Gaussian RVs $\mathcal{N}(s, \sigma^2)$ and $\mathcal{N}(0, \sigma^2)$ in the alternative ($\mathcal{H}_1$) and null ($\mathcal{H}_0$) hypotheses respectively, the distribution of the cumulative belief for a specific realization of the stopping time is given as
\vspace{-0.1cm}
\begin{align}
    (\Lambda^{(\tau^{*}_i)}_{i} | \tau^{*}_{i}, \mathcal{H}_1) \sim \mathcal{N} & \Big(\mathbb{E}[\Lambda_i^{(\tau^{*}_i)}| \tau^{*}_{i}, \mathcal{H}_1], \text{var}[\Lambda_i^{(\tau^{*}_i)}| \tau^{*}_{i} , \mathcal{H}_1]  \Big) \thinspace \thinspace \text{and},  \nonumber
\end{align}
\vspace{-0.25cm}
\begin{align}
\hspace{-0.3cm}(\Lambda^{(\tau^{*}_i)}_{i} | \tau^{*}_{i}, \mathcal{H}_0) \sim \mathcal{N} & \Big(\mathbb{E}[\Lambda_i^{(\tau^{*}_i)}| \tau^{*}_{i}, \mathcal{H}_0], \text{var}[\Lambda_i^{(\tau^{*}_i)}| \tau^{*}_{i}, \mathcal{H}_0]  \Big). \nonumber
\end{align}
where $\mathbb{E}[\Lambda_i^{(\tau^{*}_i)}| \mathcal{H}_1] = -\mathbb{E}[\Lambda_i^{(\tau^{*}_i)}| \mathcal{H}_0]= \frac{s^2 \cdot \Big(\sum_{j=1}^{\tau^{*}_{i} }w^{\tau^{*}_{i}-j}_{i}\Big )}{2 \sigma^2}$ and, $ \text{var}[\Lambda_i^{(\tau^{*}_i)}| \tau^{*}_{i} , \mathcal{H}_1] =  \text{var}[\Lambda_i^{(\tau^{*}_i)}| \tau^{*}_{i} , \mathcal{H}_0] = {s^2 \cdot \Big(\sum_{j=1}^{\tau^{*}_{i} }w^{2(\tau^{*}_{i}-j)}_{i}\Big )}/{\sigma^2}$.

\section{Simulation Results}
 In this section, we first illustrate the performance of the individual human decision-maker, who uses the decision rule as described in (\ref{human_decision}). The human decision-maker observes data from $\mathcal{N}(s, \sigma^2)$ and $\mathcal{N}(0, \sigma^2)$ under the alternative ($\mathcal{H}_1$) and null ($\mathcal{H}_0$) hypotheses respectively. We consider $s = 2$ and $\sigma^2 = 2$ for the simulations in this work. 
 
\begin{figure}[!h]
    \centering
    \includegraphics[width=55mm]{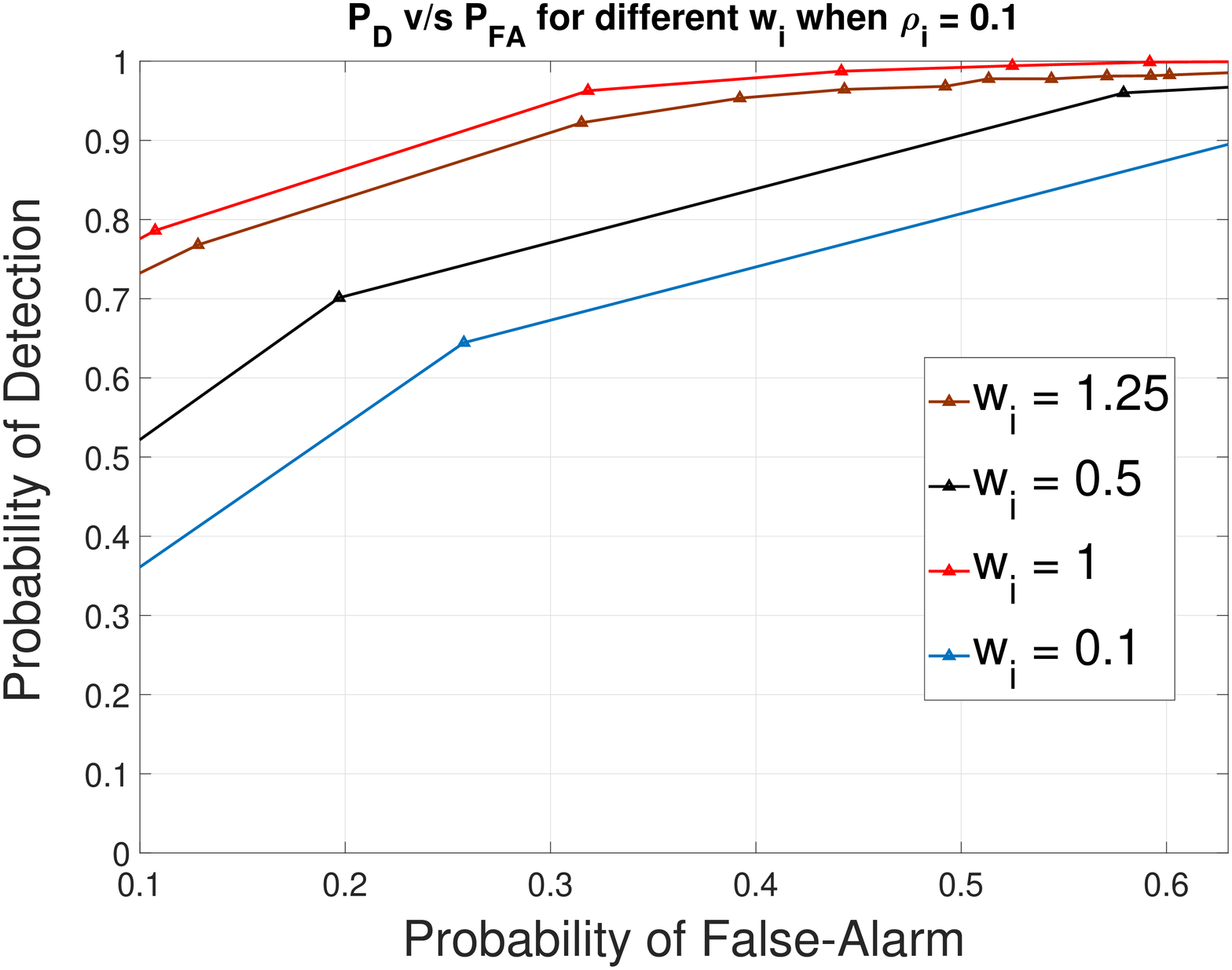}
    \caption{ROC for different values of $w_i$, when $\rho_i = 0.1$} 
    \label{fig:ROC_rho_i_0.1}
\end{figure}
\vspace{-0.1cm}
The performance of a single human with different values of $w_i$ is shown in Fig. \ref{fig:ROC_rho_i_0.1}. In this case, it can be seen that when the human decision-maker employs a random stopping time which is distributed as geometric RV with parameter $\rho_i = 0.1$, the receiver operating characteristics (ROC) curves corresponding to larger $w_i$ exhibit larger values of $P_{D,i}$ for lower values of $P_{FA,i}$. We would like to remark that the performance of a human with $w_i = 1.25$ is worse compared to a human with $w_i = 1$, which is in agreement with the illustration in Fig. \ref{fig:def_coeff_poisson} which shows that the deflection coefficient is largest when $w_i = 1$. This also validates the choice of using the deflection coefficient of the human belief LLR as a viable surrogate to measure the detection performance.

\begin{figure}[!h]
    \centering
    \includegraphics[width=59mm]{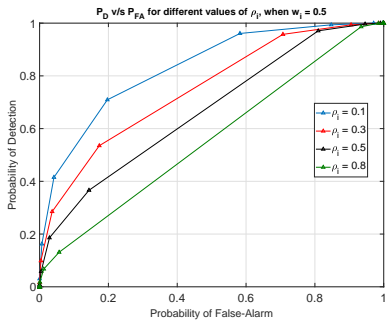}
    \caption{ROC for different values of $\rho_i$, when $w_i = 0.5$} 
    \label{fig:ROC_w_i_0.5}
\end{figure}

 The performance of a single human with different values of $\rho_i$, when $w_i = 0.5$ is shown in Fig. \ref{fig:ROC_w_i_0.5}. In Fig. \ref{fig:ROC_w_i_0.5} it can be seen that when the human decision-maker employs a random stopping time which is distributed as a geometric RV with different parameters $\rho_i = \{0.1, 0.3, 0.5, 0.8\}$, the ROC curves corresponding to smaller $\rho_i$ exhibit larger values of $P_{D,i}$ for lower values of $P_{FA,i}$. This can be explained by noting that a Geometric RV with a smaller value of $\rho_i$ is highly likely to stop at a later time, due to which a larger number of observations are used by the human, to update its cumulative belief, thereby improving the decision-making performance. 

\begin{figure}[!h]
    \centering
    \includegraphics[width=62 mm]{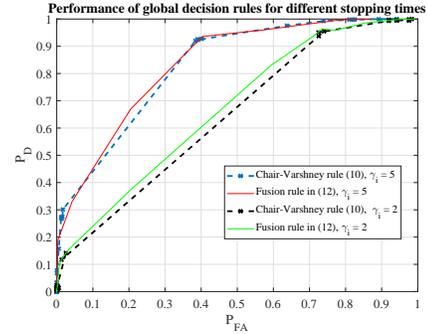}
    \caption{Performance of global decision rules with different stopping times} 
    \label{fig:global_stop_times_comparison}
\end{figure}

 \color{blue}
To illustrate the performance of the global decision rules in (\ref{global_decision_rule}) and (\ref{global_decision_rule_simplified}) we compare the performance of the decision rules in Fig. \ref{fig:global_stop_times_comparison}. 

We assume that the plots are generated with observations distributed as Gaussian $\mathcal{N}(0.5, 5)$ under $\mathcal{H}_1$ and $\mathcal{N}(0, 5)$ under $\mathcal{H}_0$. The stopping times of all the human decision-makers are distributed as a Poisson random variable with $\gamma_i = 5, \forall i =1, \dots, N$ for one set of plots and $\gamma_i = 2, \forall i =1, \dots, N$ for the other set of plots. The individual human decision thresholds are $T = [-2, 1, 1, 1, 1, 1, 1, -1, -1, -2]$ and parameters influencing the rationality of the decision-maker are $W = [1, 1, 1, 1.2, 1.25, 0.8, 0.65, 0.95, 1, 1.1]$.
\vskip 0.1cm
It can be observed from Fig. \ref{fig:global_stop_times_comparison} that the performance of the decision rules in (\ref{global_decision_rule}) and (\ref{global_decision_rule_simplified}) are similar for the same set of parameters. This demonstrates that the decision in rule in (\ref{global_decision_rule_simplified}) achieves performance close to the Chair-Varshney rule in (\ref{global_decision_rule}). Similar to the plots in Fig. (\ref{fig:ROC_w_i_0.5}) for a single human decision-maker, when the number of observations used to arrive at a decision is larger in an average sense, the performance is better. This difference can be observed in Fig. \ref{fig:global_stop_times_comparison} when the individual humans possess stopping times with different expected values, when the stopping times are distributed as Poisson random variables. 
\color{black}
\section{Conclusion}
In this work, we have considered a distributed binary hypothesis testing problem in which the decision-making agents are humans. The human decision-makers observe a sequence of i.i.d observations, which are accumulated based on a belief update model. The humans update their beliefs in such a way that they either place a larger belief on the observations that were observed earlier, or on the most recent observations. Due to different behavioral tendencies such as impatience or the inability to process observations over a long period in time, humans stop processing observations after a random point in time and perform an LLRT on the observations processed so far. We analyzed the performance of the individual human decision-maker using the deflection coefficient of human belief. We also derived the optimal fusion rule at the FC when the FC is aware of the number of random observations used by each human decision-maker for their decision.  

\clearpage
\bibliography{refer}

\begin{thebibliography}{10}
\providecommand{\url}[1]{#1}
\csname url@samestyle\endcsname
\providecommand{\newblock}{\relax}
\providecommand{\bibinfo}[2]{#2}
\providecommand{\BIBentrySTDinterwordspacing}{\spaceskip=0pt\relax}
\providecommand{\BIBentryALTinterwordstretchfactor}{4}
\providecommand{\BIBentryALTinterwordspacing}{\spaceskip=\fontdimen2\font plus
\BIBentryALTinterwordstretchfactor\fontdimen3\font minus
  \fontdimen4\font\relax}
\providecommand{\BIBforeignlanguage}[2]{{%
\expandafter\ifx\csname l@#1\endcsname\relax
\typeout{** WARNING: IEEEtran.bst: No hyphenation pattern has been}%
\typeout{** loaded for the language `#1'. Using the pattern for}%
\typeout{** the default language instead.}%
\else
\language=\csname l@#1\endcsname
\fi
#2}}
\providecommand{\BIBdecl}{\relax}
\BIBdecl

\bibitem{geng2019decision}
B.~Geng and P.~K. Varshney, ``On decision making in human-machine networks,''
  in \emph{2019 IEEE 16th International Conference on Mobile Ad Hoc and Sensor
  Systems (MASS)}.\hskip 1em plus 0.5em minus 0.4em\relax IEEE, 2019, pp.
  37--45.

\bibitem{9443353}
C.~Quan, B.~Geng, and P.~K. Varshney, ``Asymptotic performance in heterogeneous
  human-machine inference networks,'' in \emph{2020 54th Asilomar Conference on
  Signals, Systems, and Computers}, 2020, pp. 584--588.

\bibitem{Geng1911:Amelioration}
B.~Geng, P.~Varshney, and M.~Rangaswamy, ``On amelioration of human cognitive
  biases in binary decision making,'' in \emph{2019 IEEE Global Conference on
  Signal and Information Processing (GlobalSIP) (GlobalSIP 2019)}, Ottawa,
  Canada, Nov. 2019.

\bibitem{8976222}
B.~{Geng}, S.~{Brahma}, T.~{Wimalajeewa}, P.~K. {Varshney}, and
  M.~{Rangaswamy}, ``Prospect theoretic utility based human decision making in
  multi-agent systems,'' \emph{IEEE Transactions on Signal Processing},
  vol.~68, pp. 1091--1104, 2020.

\bibitem{9133140}
B.~{Geng}, Q.~{Li}, and P.~K. {Varshney}, ``Prospect theory based crowdsourcing
  for classification in the presence of spammers,'' \emph{IEEE Transactions on
  Signal Processing}, vol.~68, pp. 4083--4093, 2020.

\bibitem{sriranga2020human}
N.~Sriranga, B.~Geng, and P.~K. Varshney, ``On human assisted decision making
  for machines using correlated observations,'' in \emph{2020 54th Asilomar
  Conference on Signals, Systems, and Computers}.\hskip 1em plus 0.5em minus
  0.4em\relax IEEE, 2020, pp. 1502--1506.

\bibitem{gengheteroge}
B.~Geng, X.~Cheng, S.~Brahma, D.~Kellen, and P.~K. Varshney, ``Collaborative
  human decision making with heterogeneous agents,'' \emph{IEEE Transactions on
  Computational Social Systems}, vol.~9, no.~2, pp. 469--479, 2022.

\bibitem{geng2021utility}
B.~Geng, Q.~Li, and P.~K. Varshney, ``Utility-theory-based optimal resource
  consumption for inference in iot systems,'' \emph{IEEE Internet of Things
  Journal}, vol.~8, no.~15, pp. 12\,279--12\,288, 2021.

\bibitem{tversky1992advances}
A.~Tversky and D.~Kahneman, ``Advances in prospect theory: Cumulative
  representation of uncertainty,'' \emph{Journal of Risk and Uncertainty},
  vol.~5, no.~4, pp. 297--323, 1992.

\bibitem{mourad2016real}
S.~Mourad and A.~Tewfik, ``Real-time data selection and ordering for cognitive
  bias mitigation,'' in \emph{2016 IEEE International Conference on Acoustics,
  Speech and Signal Processing (ICASSP)}.\hskip 1em plus 0.5em minus
  0.4em\relax IEEE, 2016, pp. 4393--4397.

\bibitem{mourad2018machine}
------, ``Machine assisted human decision making,'' in \emph{2018 IEEE
  International Conference on Acoustics, Speech and Signal Processing
  (ICASSP)}.\hskip 1em plus 0.5em minus 0.4em\relax IEEE, 2018, pp. 6981--6985.

\bibitem{anderson1992discrete}
S.~P. Anderson, A.~De~Palma, and J.-F. Thisse, \emph{Discrete choice theory of
  product differentiation}.\hskip 1em plus 0.5em minus 0.4em\relax MIT press,
  1992.

\bibitem{9747866}
B.~Geng, Q.~Li, and P.~K. Varshney, ``Human decision making with bounded
  rationality,'' in \emph{ICASSP 2022 - 2022 IEEE International Conference on
  Acoustics, Speech and Signal Processing (ICASSP)}, 2022, pp. 5493--5497.

\bibitem{anderson1981foundations}
N.~H. Anderson, ``Foundations of information integration theory,'' 1981.

\bibitem{holt2009update}
C.~A. Holt and A.~M. Smith, ``An update on bayesian updating,'' \emph{Journal
  of Economic Behavior \& Organization}, vol.~69, no.~2, pp. 125--134, 2009.

\bibitem{9413745}
B.~Geng, Q.~Chen, and P.~K. Varshney, ``Cognitive memory constrained human
  decision making based on multi-source information,'' in \emph{ICASSP 2021 -
  2021 IEEE International Conference on Acoustics, Speech and Signal Processing
  (ICASSP)}, 2021, pp. 5325--5329.

\bibitem{ChairV1986}
Z.~Chair and P.~K. Varshney, ``Optimal data fusion in multiple sensor detection
  systems,'' \emph{IEEE Transactions on Aerospace and Electronic Systems}, vol.
  AES-22, no.~1, pp. 98--101, Jan. 1986.

\end{thebibliography}
\bibliographystyle{IEEEtran}

\clearpage

\section{Appendix}

\subsection{Expected Value of the Human Belief}
We first consider the case when $w_i \neq 1$. The expected value of the belief $ \Lambda^{(\tau^{*}_i)}_i$ of the $i^{th}$ human decision-maker is expanded as follows,

\begin{align}
    \mathbb{E} \Big[\Lambda^{(\tau^{*}_i)}_i \vert \mathcal{H}_k \Big] =& \quad \mathbb{E} \bigg[ \sum_{j=1}^{\tau^{*}_{i}} \thinspace w_{i}^{\tau^{*}_{i} - j}\lambda^{(j)}_i \bigg \vert \mathcal{H}_k  \Bigg] \nonumber\\
    \overset{(a)}{=}&  \quad \mathbb{E}_{\tau^{*}_{i}} \Bigg \{ \mathbb{E} \Bigg[ \sum_{j=1}^{\tau^{*}_{i}} \thinspace w_{i}^{\tau^{*}_{i} - j}\lambda^{(j)}_i \bigg \vert \mathcal{H}_k, \tau^{*}_{i}  \Bigg] \Bigg \} \nonumber\\
    \overset{(b)}{=}&  \quad \mathbb{E}_{\tau^{*}_{i}} \Bigg \{ \sum_{j=1}^{\tau^{*}_{i}} \thinspace \Big(w_{i}^{\tau^{*}_{i} - j} \Big) \cdot \mathbb{E} \Big[\lambda^{(j)}_i \Big \vert \mathcal{H}_k \Big] \bigg \vert \thinspace \tau^{*}_{i}  \Bigg \} \nonumber\\
    \overset{(c)}{=}&  \quad \mathbb{E} \Big[\lambda^{(j)}_i \Big \vert \mathcal{H}_k \Big] \cdot  \mathbb{E}_{\tau^{*}_{i}} \Bigg [ \sum_{j=1}^{\tau^{*}_{i}} \thinspace \Big(w_{i}^{\tau^{*}_{i} - j} \Big)   \Bigg ] \nonumber\\ 
    \overset{(d)}{=}&  \quad \mathbb{E} \Big[\lambda^{(j)}_i \Big \vert \mathcal{H}_k \Big] \cdot  \mathbb{E}_{\tau^{*}_{i}} \Bigg [ \frac{w_{i}^{\tau_{i}^{*}}-1}{w_i - 1}  \Bigg ] \nonumber\\
    \overset{(e)}{=}&  \quad \mathbb{E} \Big[\lambda^{(j)}_i \Big \vert \mathcal{H}_k \Big] \cdot   \Bigg [ \frac{\mathbb{E}_{\tau^{*}_{i}}[w_{i}^{\tau_{i}^{*}}]-1}{w_i - 1}  \Bigg ] \nonumber\\
    \label{eq:expected_LLR_belief}
\end{align}
where (a) is due to the law of total expectation, (b) is due to the linearity property of the expectation operator, (c) is due to the fact that $\tau^{*}_{i}$ is independent of $\Lambda^{(\tau^{*}_{i})}_{i}$, (d) is obtained by summing the terms $w^{\tau_{i}^{*}-j}_{i}$ which constitute a geometric progression and (e) is obtained by moving the expectation into the argument. 

The quantity $\mathbb{E}_{\tau^{*}_{i}} \Big[w_{i}^{\tau_{i}^{*}} \Big] = \mathbb{E}_{\tau^{*}_{i}} \Big[e^{ \text{ln}(w_i) \thinspace \tau_{i}^{*}} \Big]$ is the moment-generating function (MGF) of the random variable $\tau_{i}^{*}$, evaluated at $\text{ln}(w_i)$ where $\text{ln}(\cdot)$ is the natural logarithm. The product term in (\ref{eq:expected_LLR_belief}) is evaluated as
\begin{align}
     \frac{\mathbb{E}_{\tau^{*}_{i}}[w_{i}^{\tau_{i}^{*}}]-1}{w_i - 1}  = 
     \begin{cases}
        \mathbb{E}_{\tau^{*}_{i}}[\tau_{i}^{*}]  & \text{if } w_i = 1\\
        \frac{\mathbb{E}_{\tau^{*}_{i}}\Big[e^{ \text{ln}(w_i) \thinspace \tau_{i}^{*}} \Big]-1}{w_i - 1} & \text{otherwise},  
    \end{cases}
    \label{eq:expected_time_RV}
\end{align}
which can be obtained by directly substituting $w_i$ with 1, in the expression in (c) of (\ref{eq:expected_LLR_belief}).

\subsection{Variance of the human belief}
The variance of the human belief LLR can also be analyzed by followoing a method similar to that of the expected value as follows, 

\begin{align}
    \text{var} \Big[\Lambda^{(\tau^{*}_i)}_i \vert \mathcal{H}_k \Big] =& \quad \text{var} \bigg[ \sum_{j=0}^{\tau^{*}_{i}} \thinspace w_{i}^{\tau^{*}_{i} - j}\lambda^{(j)}_i \bigg \vert \mathcal{H}_k  \Bigg] \nonumber\\
    \overset{(a)}{=}&  \quad \mathbb{E}_{\tau^{*}_{i}} \Bigg \{ \text{var} \Bigg[ \sum_{j=0}^{\tau^{*}_{i}} \thinspace w_{i}^{\tau^{*}_{i} - j}\lambda^{(j)}_i \bigg \vert \mathcal{H}_k, \tau^{*}_{i}  \Bigg] \Bigg \}  \nonumber\\
    &+ \text{var}_{\tau^{*}_{i}} \Bigg \{ \mathbb{E} \Bigg[ \sum_{j=0}^{\tau^{*}_{i}} \thinspace w_{i}^{\tau^{*}_{i} - j}\lambda^{(j)}_i \bigg \vert \mathcal{H}_k, \tau^{*}_{i}  \Bigg] \Bigg \} \nonumber\\
    \overset{(b)}{=}&  \quad \mathbb{E}_{\tau^{*}_{i}} \Bigg \{ \sum_{j=0}^{\tau^{*}_{i}} \thinspace \Big(w_{i}^{\tau^{*}_{i} - j} \Big)^2 \cdot \text{var} \Big[\lambda^{(j)}_i \Big \vert \mathcal{H}_k \Big] \bigg \vert \thinspace \tau^{*}_{i}  \Bigg \} \nonumber\\
    &+ \text{var}_{\tau^{*}_{i}} \Bigg \{ \Bigg[ \sum_{j=0}^{\tau^{*}_{i}} \thinspace w_{i}^{\tau^{*}_{i} - j} \cdot  \mathbb{E}\Big [\lambda^{(j)}_i \bigg \vert \mathcal{H}_k \Big] \bigg \vert  \tau^{*}_{i}  \Bigg] \Bigg \} \nonumber\\
    \overset{(c)}{=}&  \quad \text{var} \Big[\lambda^{(j)}_i \Big \vert \mathcal{H}_k \Big] \cdot  \mathbb{E}_{\tau^{*}_{i}} \Bigg [ \sum_{j=0}^{\tau^{*}_{i}} \thinspace \Big(w_{i}^{\tau^{*}_{i} - j} \Big)^2   \Bigg ] \nonumber\\ 
    &+ \mathbb{E}\Big [\lambda^{(j)}_i \bigg \vert \mathcal{H}_k \Big] \cdot \text{var}_{\tau^{*}_{i}} \Bigg \{ \Bigg[ \sum_{j=0}^{\tau^{*}_{i}} \thinspace w_{i}^{\tau^{*}_{i} - j} \bigg \vert  \tau^{*}_{i}  \Bigg] \Bigg \} \nonumber\\
    \overset{(d)}{=}&  \quad \text{var} \Big[\lambda^{(j)}_i \Big \vert \mathcal{H}_k \Big] \cdot  \mathbb{E}_{\tau^{*}_{i}} \Bigg [ \frac{w_{i}^{2\cdot \tau_{i}^{*}}-1}{w^{2}_i - 1}  \Bigg ] \nonumber\\
    &+ \mathbb{E}\Big [\lambda^{(j)}_i \bigg \vert \mathcal{H}_k \Big] \cdot \text{var}_{\tau^{*}_{i}} \Bigg [ \frac{w_{i}^{\tau_{i}^{*}}-1}{w_i - 1} \Bigg ] \nonumber\\
    \overset{(e)}{=}&  \quad \text{var} \Big[\lambda^{(j)}_i \Big \vert \mathcal{H}_k \Big] \cdot   \Bigg [ \frac{\mathbb{E}_{\tau^{*}_{i}}[w_{i}^{2 \cdot \tau_{i}^{*}}]-1}{w^2_i - 1}  \Bigg ] \nonumber\\
    &+ \mathbb{E}\Big [\lambda^{(j)}_i \bigg \vert \mathcal{H}_k \Big] \cdot  \Bigg [ \frac{\text{var}_{\tau^{*}_{i}}[w_{i}^{\tau_{i}^{*}}]}{(w_i - 1)^2} \Bigg ]
    \label{eq:var_LLR_belief}
\end{align}
where (a) is due to the law of total variance, (b) is due to the linearity property of the expectation operator in the second term and due to the fact that $\lambda^{(j)}_{i}$s are independent of each other, the variance of the sum of terms is equal to the sum of variances. Equation (c) is due to the fact that $\tau^{*}_{i}$ is independent of $\Lambda^{(\tau^{*}_{i})}_{i}$, (d) is obtained by summing the terms $w^{\tau_{i}^{*}-j}_{i}$ and $w^{2 \cdot \tau_{i}^{*}-j}_{i}$ which are terms corresponding to geometric progressions and (e) is obtained by moving the expectation and variance into their respective arguments.

\section{Additional plots}

\begin{figure}[!h]
    \centering
    \includegraphics[width=50mm]{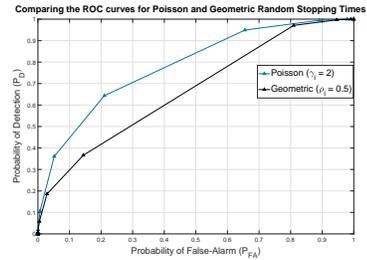}
    \caption{Comparing the ROCs of Poisson and Geometric stopping times with equal means} 
    \label{fig:ROC_Geo_Poi_w_i_0.5}
\end{figure}

In Fig. \ref{fig:ROC_Geo_Poi_w_i_0.5}, we compare the decision-making performance of a human when the human employs a Geometric random stopping time and a Poisson random stopping time, respectively. The parameter $w_i$ is set to $0.5$ for both cases. The parameter $\rho_i = 0.1$ and the parameter $\gamma_i = 2$. It is to be noted that the expected value of the stopping times for these choices of parameters is such that the random stopping times have the same expected value. However, the ROC has a larger area under the curve when the random stopping time is a Poisson RV in contrast with the case when the random stopping time is a Geometric RV. 

\end{document}